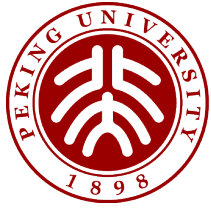

# IMPARTIAL REDISTRICTING: A MARKOV CHAIN APPROACH TO THE "GERRYMANDERING PROBLEM"

A THESIS

SUBMITTED TO THE PROGRAM IN COMPUTER SCIENCE OF PEKING UNIVERSITY IN FULFILLMENT OF THE REQUIREMENTS FOR THE DEGREE OF BACHELORS OF SCIENCE

Jason Xiaotian Dou

June 2014

**Thesis Evaluation Form**

Name: Jason Xiaotian Dou    Title of Thesis: Impartial Redistricting: A Markov Chain Approach to The "Gerrymandering Problem"

University: Peking University

| Evaluation Criteria | Grade |
|---|---|
| **Problem definition**<br>- relevant<br>- clearly phrased<br>- testable | A |
| **Execution**<br>- scholarly level<br>- level of innovation | A |
| **Analysis, interpretation, conclusions**<br>- clear<br>- defendable | A |
| **Clearly phrased reporting** | A |
| **Structure of the thesis** | A |
| **Further comments (Cons and Pros)** | |

| The project turned out to be more difficult than I first envisioned. Jason stuck to the task and | made a good first start. |

Signature *Alam Fnesi*

# Thesis Evaluation Form

Name: Jason Xiaotian Dou  Title of Thesis: Impartial Redistricting: A Markov Chain Approach to The "Gerrymandering Problem"

University: Peking University

| Evaluation Criteria | Grade |
|---|---|
| **Problem definition** | |
| • relevant | A |
| • clearly phrased | A |
| • testable | A |
| **Execution** | |
| • scholarly level | B |
| • level of innovation | B |
| **Analysis, interpretation, conclusions** | |
| • clear | A |
| • defendable | B |
| **Clearly phrased reporting** | A |
| **Structure of the thesis** | A |
| **Further comments (Cons and Pros)** | The reason I have given him a few "B"s above is because there are a few more things that I wish he had had time to do. But he did a nice job on the parts that he completed. |

Signature  *Daniel Sleator*  June 5, 2014
Professor of Computer Science
Carnegie Mellon University

# Abstract

After every U.S. national census, a state legislature is required to redraw the boundaries of congressional districts in order to account for changes in population. At the moment this is done in a highly partisan way, with districting done in order to maximize the benefits to the party in power. This is a threat to U.S's democracy.

There have been proposals to take the re-districting out of the hands of political parties and give to an "independent" commission. Independence is hard to come by and in this thesis we want to explore the possibility of computer generated districts that as far as possible to avoid partisan "gerrymandering".

The idea we have is to treat every possible redistricting as a state in a Markov Chain: every state is obtained by its former state in random way. With some technical conditions, we will get a near uniform member of the states after running sufficiently long time (the mixing time). Then we can say the uniform member is an impartial distribution.

Based on the geographical and statistical data of Pennsylvania, I have achieved the Markov Chain algorithm with several constraints, done optimization experiments and a web interface is going to be made to show the results.

# Acknowledgments

First of all, thanks to my parents. For doing all the essential things those parents do (and being the best in the world in my mind). Particularly they offer me great space to grow up freely, encourage me to think big, act small.

Thanks to my thesis advisors Professor Danny Sleator in Computer Science Department, Alan Frieze in Mathematics Department and Professor David Miller in Department of History at Carnegie Mellon University (CMU), United States of America. The thesis is mainly based on the research project I have done in summer, 2013 with them at Carnegie Mellon. I am so fortunate to have the chance being with them. Especially thanks to Professor Danny Sleator, though he looks a little tough at first glance, he indeed spent lots of time guiding me, listening to me and discussing with me on the algorithmic part of the project in the lovely CMU-ACM training room in Gates Building. And he is also good at teaching me English and the determination as a "Climate Hawk". Thanks to Professor Alan frieze, he taught me the secret of Markov Chain humorously and has precious sense of the whole picture of the project. Hope we can finally publish the project on the Science Magazine as he said. Thanks to Professor David Miller, it was he that initiated the project and did lots of data analysis work to provide geographical and statistical data for me.

Thanks to my local advisor Professor Junfeng Hu at Institute of Computational Linguistics, Peking University. I recognized him in my sophomore year. It is his guide and encouragement that make me fall in love with computer science and be confident to handle whatever difficulties I face during my work. Also thanks to his work on allowing me to do the CMU project as my thesis and kind remainder during the process.

Thanks to Associate Dean Wenxin Li, Professor Hui Zhang, Professor Yao Guo, Mrs Zhaohui Yang and Xiang Li for support of the PKU-CMU research program. Thanks to Professor Fan Ye, Mingmin Zhao, Zhiting Hu, Jason Li for all the interesting discussions, insightful questions and ideas.

# Contents

## 1 Preliminaries

### 1.1 Markov Chain and Mixing Time

A Markov chain is a sequence of random variables X1, X2, X3, ... with the Markov property, namely that, given the present state, the future and past states are independent. Formally,

$$\Pr(X_{n+1} = x \mid X_1 = x_1, X_2 = x_2, \ldots, X_n = x_n) = \Pr(X_{n+1} = x \mid X_n = x_n)$$

, if both sides of the equation are well defined. The possible values of Xi form a countable set S called the state space of the chain. Markov chains are often described by a directed graph, where the edges are labeled by the probabilities of going from one state to the other states.

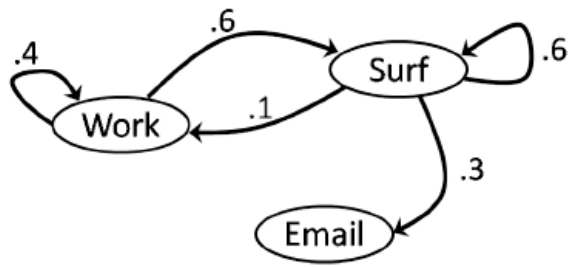

Figure 1.1

In probability theory, the mixing time of a Markov chain is the time until the Markov chain is "close" to its steady state distribution.

More precisely, a fundamental result about Markov chains is that a finite state irreducible aperiodic chain has a unique stationary distribution π and, regardless of the initial state, the time-t distribution of the chain converges to π as t tends to infinity. Mixing time refers to any of several variant formalizations of the idea: how large must t be until the time-t distribution is approximately π? One variant, variation distance mixing time, is defined as the smallest t such that

$$|\Pr(X_t \in A) - \pi(A)| \leq 1/4$$

## 1.2 Metropolis chain

The Metropolis–Hastings algorithm can draw samples from any probability distribution P(x), provided you can compute the value of a function f(x) which is proportional to the density of P. The lax requirement that f (x) should be merely proportional to the density, rather than exactly equal to it, makes the Metropolis–Hastings algorithm particularly useful, because calculating the necessary normalization factor is often extremely difficult in practice.

## 1.3 The data issues

A Census County Division (CCD) is a subdivision of a county used by the United States Census Bureau for the purpose of presenting statistical data. A CCD is a relatively permanent statistical area delineated cooperatively by the Census Bureau and state and local government authorities.

```
PA2010CtySubDiv.txt
 1  _SHAPE_TAG_=Area
 2  STATEFP10=42
 3  COUNTYFP10=003
 4  COUSUBFP10=23616
 5  COUSUBNS10=01214794
 6  GEOID10=4200323616
 7  NAME10=Emsworth
 8  NAMELSAD10=Emsworth borough
 9  LSAD10=21
10  CLASSFP10=C5
11  MTFCC10=G4040
12  CNECTAFP10=
13  NECTAFP10=
14  NCTADVFP10=
15  FUNCSTAT10=F
16  ALAND10=1472769
17  AWATER10=318264
18  INTPTLAT10=+40.5108721
19  INTPTLON10=-080.0962166
20  -211319.7526198384600;-51228.3783962036470
21  -211331.9314629652100;-51221.0315229659540
22  -211388.6684374962100;-51187.3043416792720
23  -211491.7867844421700;-51134.7023501110740
24  -211651.3531260339700;-51053.2683000836260
25  -211750.1950393886100;-51002.5620333110640
26  -211846.6936209474200;-50953.0323139453030
27  -211887.9963923805600;-50931.6317429697300
28  -212132.7160068774200;-50804.8672291450860
29  -212207.1249647835800;-50769.2977732707950
30  -212210.7664955897800;-50697.3909179857950
31  -212215.5517842783300;-50588.1048754899820
32  -212216.1001433302000;-50580.5310412990580
33  -212218.1370681203800;-50553.6858247835770
34  -212218.9361403797500;-50542.9926436450330
35  -212219.0646560080300;-50541.5440317780960
36  -212219.7981111542000;-50510.6232573518570
37  -212219.7598322485300;-50506.2894610646620
38  -212220.1628342200700;-50469.7093771408690
39  -212177.8546955248500;-50389.8821753589360
```

Figure 1.2

The tract-based data is called "census tracts" of the 2010 census by the census bureau. Many of the tracts are identical to the county subdivisions as used before. The tracts, however, are used at the very earliest stage of the census decade process, and the only published items of demographic tract data are the number of people living in the tract and the number of houses. Because most social scientists are interested in more detailed social information (age, income, occupation, education, sex, race, etc.) researchers probably use tracts much less frequently than county subdivisions or any of several other census bureau geographic divisions. However, since our project requires polygonal divisions geographically and needs only the total population demographically, the tracts appear to be our best choice of divisions. Significantly there are fewer "composites" and "groups" in the tracts than there are in the county subdivisions.

# 2 Introduction

## 2.1 The Gerrymandering Problem

In the process of setting electoral districts, gerrymandering is a practice that attempts to establish a political advantage for a particular party or group by manipulating district boundaries to create partisan advantaged districts. The resulting district is known as a gerrymander; however, that word can also refer to the process. When used to allege that a given party is gaining disproportionate power, the term gerrymandering has negative connotations.

In addition to its use achieving desired electoral results for a particular party, gerrymandering may be used to help or hinder a particular demographic, such as a political, ethnic, racial, linguistic, religious, or class group, such as in U.S. federal voting district boundaries that produce a majority of constituents representative of African-American or other racial minorities, known as "majority-minority districts".

## 2.2  The Gerrymandering Voting results at State of Pennsylvania

Let we use the State of Pennsylvania as an example, in the House of Representatives, there are 18 representatives for the 18 districts in Pennsylvania, then people in one district vote for the representative in their districts. Figure 1.1 shows the result of 2010 voting at Pennsylvania, the Republican as the party in power at the state gained only 48.8% of votes but 72.2% of seats at the House of Representatives. But how do they achieve this? Figure 1.2 shows the districting results drawn by Republicans. The shape is quite "Gerrymandering" indeed.

| Congressional district | Republican votes | Democratic votes | other votes | total votes | winner |
|---|---|---|---|---|---|
| 1st | 41,708 | 235,394 |  | 277,102 | Democrat |
| 2nd | 33,381 | 318,176 | 4,829 | 356,386 | Democrat |
| 3rd | 165,826 | 123,933 | 12,755 | 302,514 | Republican |
| 4th | 181,603 | 104,643 | 17,734 | 303,980 | Republican |
| 5th | 177,740 | 104,725 |  | 282,465 | Republican |
| 6th | 191,725 | 143,803 |  | 335,528 | Republican |
| 7th | 209,942 | 143,509 |  | 353,451 | Republican |
| 8th | 199,379 | 152,859 |  | 352,238 | Republican |
| 9th | 169,177 | 105,128 |  | 274,305 | Republican |
| 10th | 179,563 | 94,227 |  | 273,790 | Republican |
| 11th | 166,967 | 118,231 |  | 285,198 | Republican |
| 12th | 175,352 | 163,589 |  | 338,941 | Republican |
| 13th | 93,918 | 209,901 |  | 303,819 | Democrat |
| 14th | 75,702 | 251,932 |  | 327,634 | Democrat |
| 15th | 168,960 | 128,764 |  | 297,724 | Republican |
| 16th | 156,192 | 111,185 | 17,404 | 284,781 | Republican |
| 17th | 106,208 | 161,393 |  | 267,601 | Democrat |
| 18th | 216,727 | 122,146 |  | 338,873 | Republican |
| total | 2,710,070 | 2,793,538 | 52,722 | 5,556,330 |  |
|  |  |  |  |  |  |
| % OF VOTES | 48.8% | 50.3% | 0.9% |  |  |
| % OF SEATS | 72.2% | 27.8% | 0.0% |  |  |

Figure 2.1

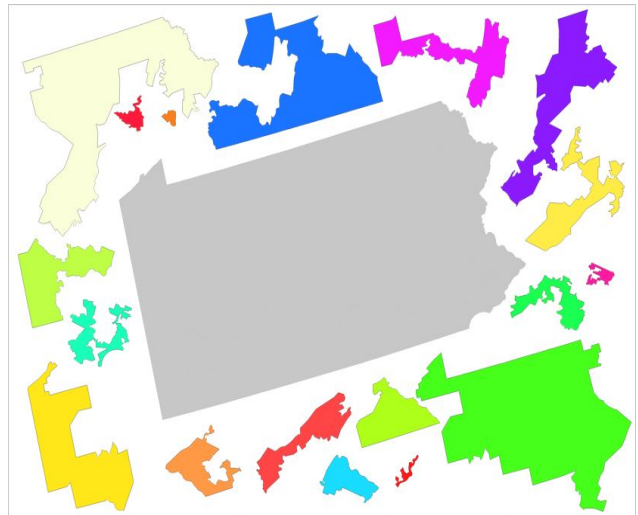

Figure 2.2

## 3 Related Works

People has proposed and applied many different methods on "Gerrymandering Problem" in general so far, however as far as we know, there has been no work working on the State of Pennsylvania yet. Works have been down including solving the districting problem by modeling it as a Markov decision process rewarding traditional measures of district "goodness": equality of population, continuity, preservation of county lines, and compactness of shape. And Multi-Seeded Growth Model simulates the creation of a fixed number of districts for an arbitrary geography by "planting seeds" for districts and specifying particular growth rules. There is also work using simulated annealing algorithm to find the simplest and fairest way to draw districts on a state.

## 4 Ideas and The Structure of Modeling Methods

There are 2518 County subdivisions (around 3000 tracts instead) at Pennsylvania, which are the basic units made up of the 18 districts. What's more, the districts need to be simply connected, compact and balanced in Population. Referring to compact, we define a compact score to measure compactness of districting results as the following:

$$\text{compact score} = \sum_{i=1}^{18} \sqrt{\text{area}_i} \Big/ \text{circumference}_i$$

Referring to "balanced in population", we use the balanced score, which is population's standard error divide mean to quantify:

$$\text{balanced score} = \frac{18 * \sqrt{\sum_{i=1}^{18}(pi - \bar{p})^2}}{17 * \sum_{i=1}^{18} pi}$$

### 4.1 Definition of "Impartial Redistricting"

"Impartial" or "fair" is quite a subjective requirement indeed. I proposed two ways for definition of "Impartial" here. The first one is that reasonable districting will have the same chance to appear. Then we need a criterion for "reasonable". This includes simply connected, "compact" and "balanced in population". And then how to get every districting has the same chance to appear? We need to use Metropolis Algorithm to modify the final stationary distribution into a uniform distribution. So every "reasonable districting" has the same probability to appear. The other way for "Impartial Districting" is that the better the districting is, the higher probability the districting will have to appear. "Better" will also be a mixture of simply connected, balanced in population and compact. This will also be achieved by the metropolis algorithm.

### 4.2 Problem abstraction

Think of a state as divided into precincts P1, P2,..., PM where precinct Pi，has eligible voting population pi for i = 1,2,...,M. The set [M] = {1,2, ..., M} must be partitioned into districts D = (D1, D2, . . . , DK ).

There are various requirements that D must satisfy. The most obvious is that the populations Π j = Σi∈Dj pi must be "close" in size. Another is geometric. The area covered by each Dj should be connected (ignoring questions about islands in the middle of large rivers).

The results of re-districting can be highly unrepresentative. As an artificial example suppose that 3 divides M and that M = 100K. There are two parties A and B and A does the re-districting. Suppose that all precints have the same population and p1, p2,..., pM/3 vote solidly for party B, whereas in the remaining districts the vote will be 51% for party A and 49% for party B. Suppose now we make a partition Di = {100(i − 1) + 1,...,100i} for i = 1,2,...,K. Then in an election party A will win 2/3 of the seats while party B will receive 66% of the votes cast.

## 5 Detailed Algorithmic Work

## 5.1 the "dual graph"

How do we model the geographical data in State of Pennsylvania in our algorithm? The way Professor Danny Sleator proposed is that we model every "county-subdivision" ("tract" in the new version of data) as a simple polygon. The first advantage is for efficient computing. The geographical data points, which consist of the boundaries of basic districting unit, are quite dense and most of them are redundant in our algorithm at all (though redundant in the Markov Chain Algorithm, they are necessary in the visualization part). So we only pick up the crucial points (which are at the corner between different districting unit). The method we are going to pick up the points is to test whether they have more than or equal to three unit neighbors.

The second method is to take the "Dual Graph" of the first method. In order to decrease the reliability of the boundary data, instead of modeling the basic unit of districting as a polygon, in this case we model it as a point in graph. The attributes like population, circumferences and area of the unit are saved as attribute in the point. And we need to calculate adjacent information advance from the boundary data to extract the adjacent information between units. The first method was completely implemented in java language with map visualization. The rest of this part will write on the details. The second method is presented as a new algorithm proposal for further experiments.

## 5.2 Random Walk Algorithm

In our model, every redistricting is treated as a state in a Markov Chain, then state transition is achieved by randomly flipping basic units – county subdivisions. Every transition, the simply connected function is used to reject "not simply connected" movement, and functions controlling compactness and balanced in population is also achieved.

| Algorithm: Markov Chain to Redistricting |
|---|
| Data preprocessing and graph model built |
| While(n--){    // n is the iteration times, that is the random walk times in Markov Chain |
|     Randomly pick up one boundary line between two districts; |
|     Randomly choose one boundary county-subdivision next to the boundary line; |
|     If(flipping will not obey simply connected and other constraints) |
|       Flip the county-subdivision from the original district to the adjacent districts; |
| } |

## 5.3 The Initial State and "Prerun Mode"

According to the memory less property of Markov Chain, initial state will not determine the ultimate results. So we build an initial state like this, there are 17 districts containing only 1 county-subdivision and other 1 district containing the rest of all the county subdivision. An illustration of it on 50 * 50 grid test data is like Figure 3.1. It is called initial state. Then a "prerun" mode is implemented to grow the small district into bigger one. Only slightly change is needed from the former main algorithm to achieve this mode.

| Algorithm: "Prerun" Mode |
|---|
| Data preprocessing and graph model built |
| While(n--){    // n is the iteration times, that is the random walk times in Markov Chain |

> Randomly pick up one boundary line between two districts;
> Randomly choose one boundary county-subdivision next to the boundary line;
> If(flipping will not obey simply connected and other constraints & the original district is bigger than the adjacent districts(regarding to the number of county subdivisions))
> Flip the county-subdivision from the original district to the adjacent districts;
> }

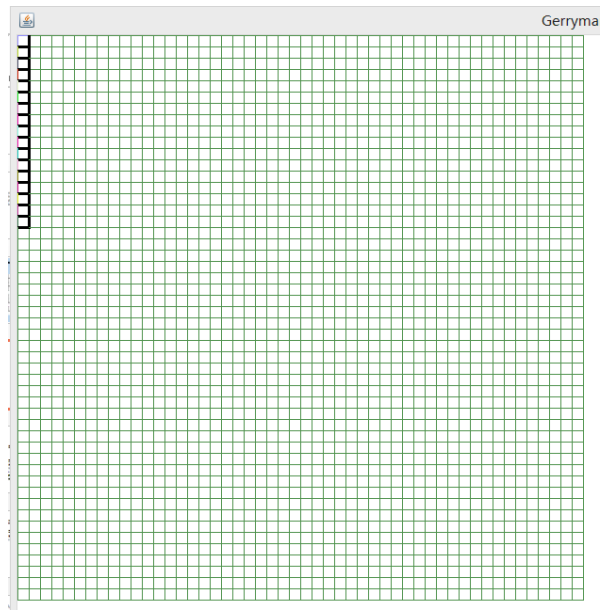

Figure 5.1

## 5.4 Second Modeling Method

Before we abstracted the tract as polygon, and pick up boundary lines to do random walk, while this relies on precise boundary data of tract which we don't have. Now instead I hope to abstract the tract as one point, not consider their geographical shape during random walk. But just paint their shapes when visualization. Since during the random walk, the constraint regarding to the geographical shape is only "simply-connected". Maybe adjacency relationship between tracts is enough for achieving it.

So we assume from Professor David Miller's data, we can get precise adjacency relationship information (by "adjacency relationship" I mean tract Tom is adjacent to tract Jerry). Since there are around 2000 tracts. So we keep a 2000 * 2000 0-1 matrix for the adjacent relationship.

Then we check the simply connected constraints only based on the matrix. This will make the algorithm amazingly easy.

# 6  Explore the Geographical Data

There are three kinds of county-subdivision type of the state of Pennsylvania shows as Figure 6.1: simple polygon, composite object and group of objects. Different computational geometry methods are applied to deal with them.

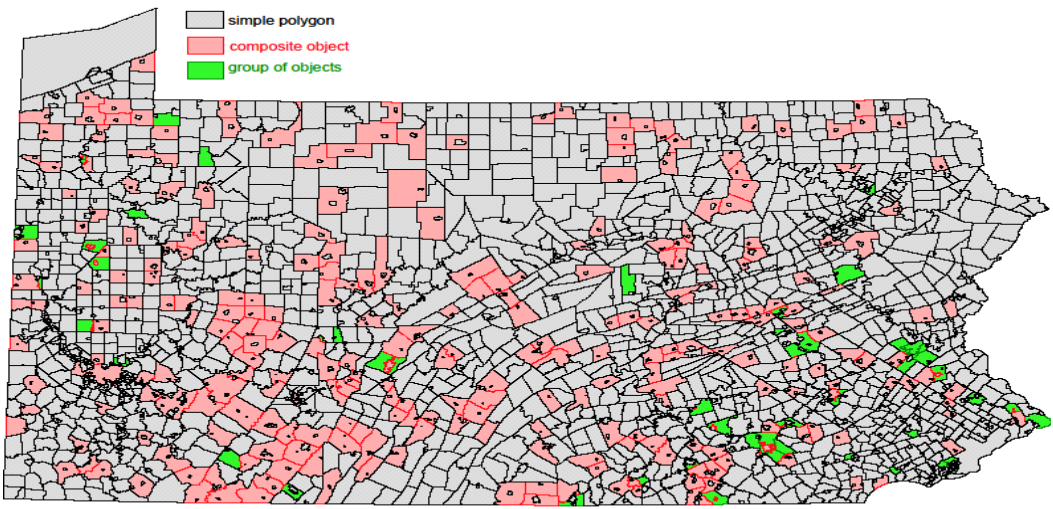

Figure 6.1

There are also many different kinds of problems in the data format as the following figures show.

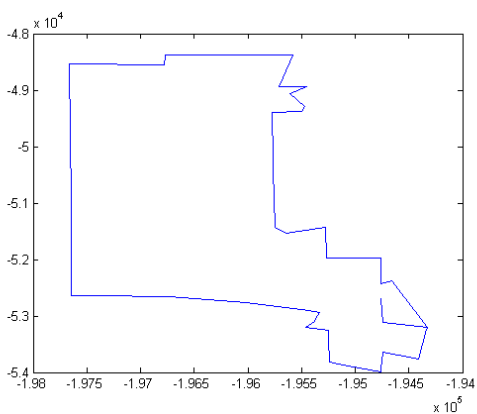
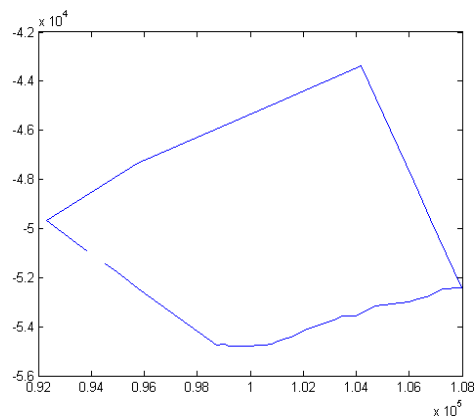

Figure 6.2                                              Figure 6.3

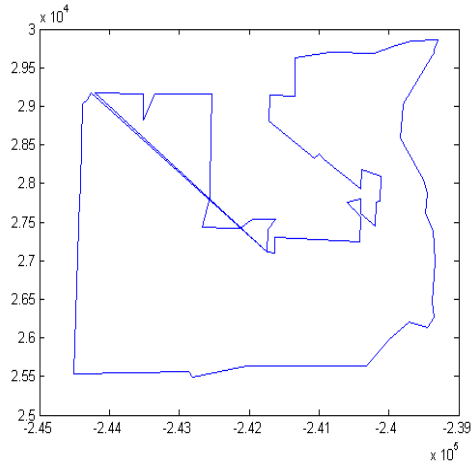

Figure 6.4

# 7   Results and Discussion

Figure 7.1 is a districting result on the 50 * 50 test data, different colors enclosed by black lines are different districts, every small grid denotes one county-subdivisions. The result is after 40000 times "prerun" mode and 20000 times "run" model. This is a result with only "simply connected" control.

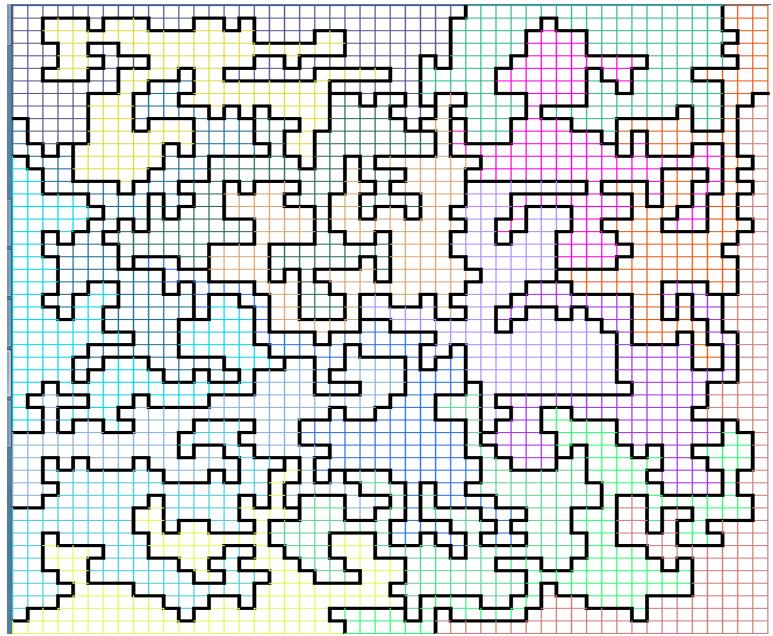

Figure 7.1

Figure 7.2 shows districting result after 50000 random walk on the county-subdivision based data. Different colors denote different districts. Since there are many problems with the data, so the districting result is not very desirable yet.

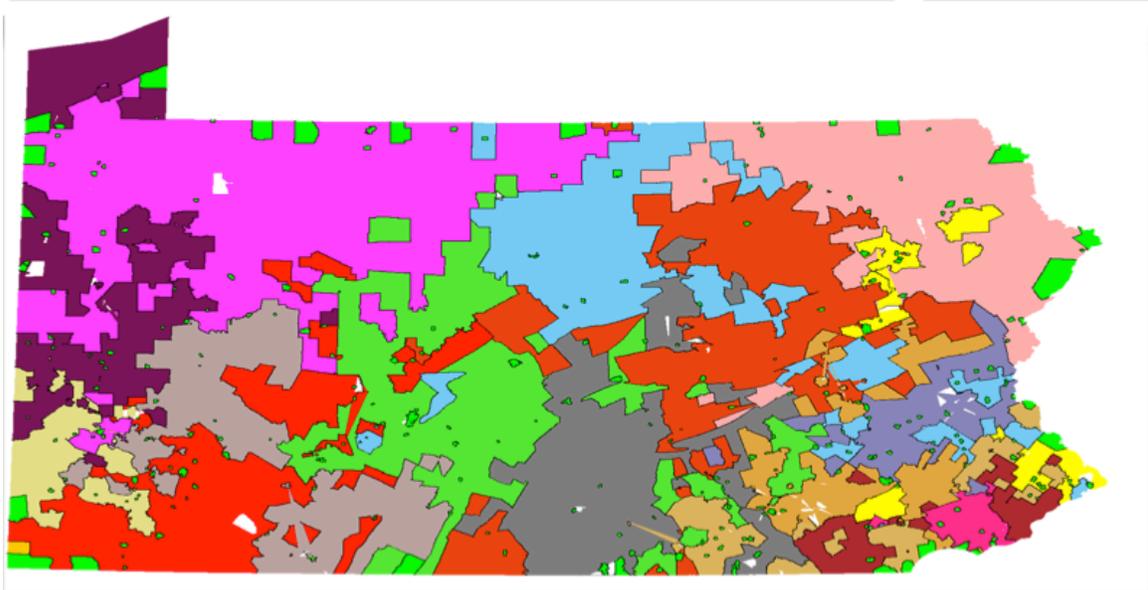

Figure 7.2

Figure 7.3 shows the balanced score's variation during four times 40000 + 20000 random walk. Figure 7.4 is the average of the four lines on Figure 7.3. It is obviously that there is an increasing trend of the balanced score; we can understand is as that entropy of the system increase during the random walk process.

To gain knowledge on how to get "compact" and "balanced in population" districting, a population following normal distribution (mean 100, standard variance 50) is assigned to every county subdivision. The experimental results of the baseline and simulated-annealing are showed in Figure 7.3, Figure 7.4 and Figure 7.5.

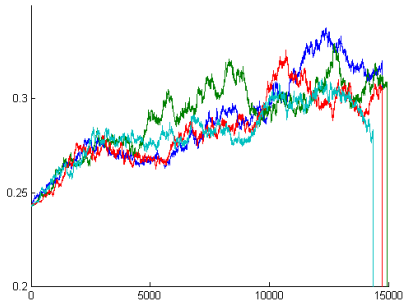 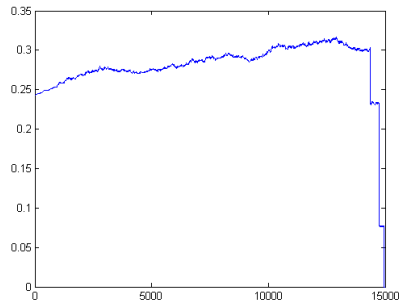

Figure 7.3                               Figure 7.4

Figure 7.5 shows the balanced score's variation during 40000 + 20000 times' random walk after applying simulated-annealing optimization. Four times of experiments are showed in the figure. It seems there are two groups of cases and many interesting phenomenon is hidden.

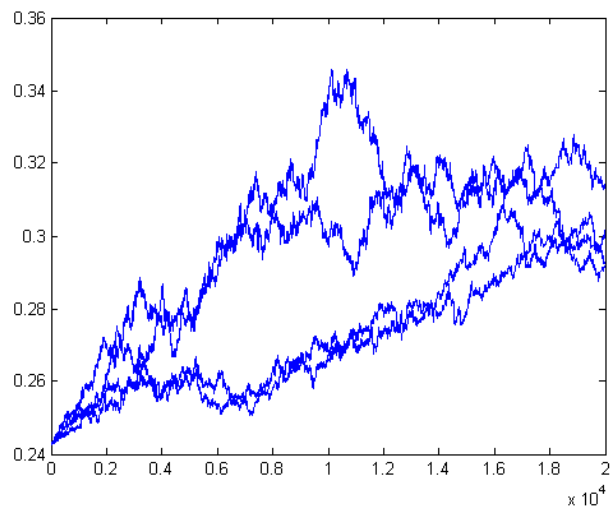

Figure 7.5

This additional part summarizes new result of redistricting, Figure 7.6 shows the initial state, Figure 7.7 shows the result after pre-run mode 200000 times and a total compact score of 3.0, and Figure 7.8 shows result with a total compact score of 3.0 and a district's own threshold of 0.11.

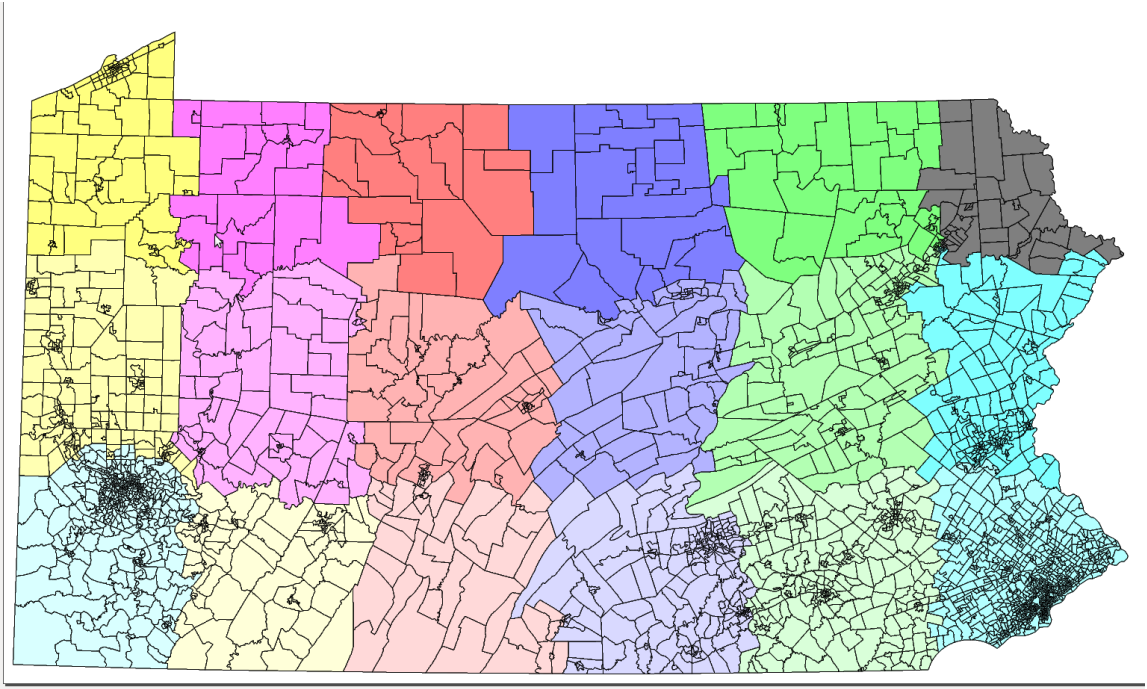

Figure 7.6

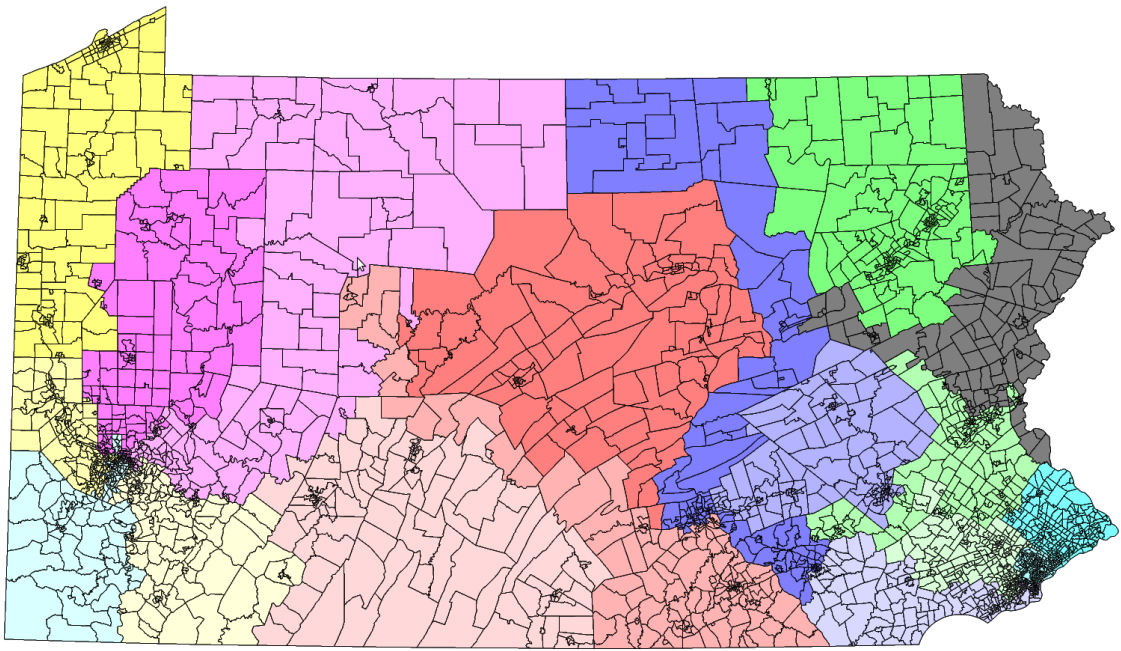

Figure 7.7

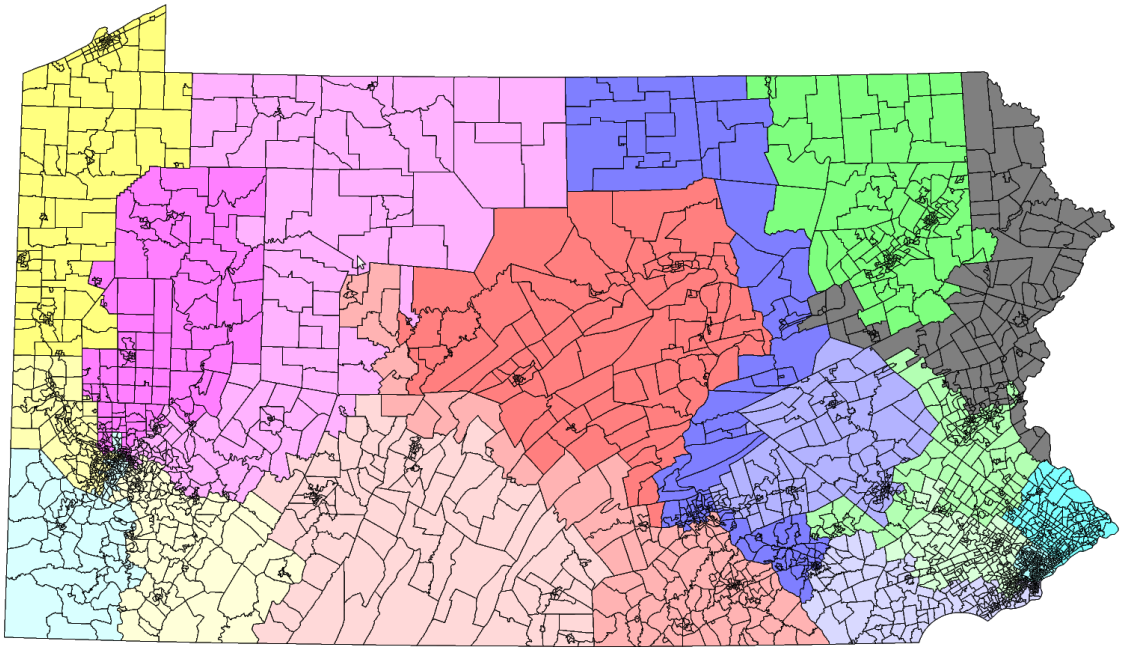

Figure 7.8

## 8 Conclusion

Random walk on the real data is much more complicated than 50 * 50 test grid data, before we abstract every subdivision as a simple polygon, however, this rely on the perfectness of the data, which is hard to achieve after several tries. A new abstraction method is proposed to decrease the reliability of data, which is to treat every county-subdivision as a node in the graph, then save information regarding population, boundary length and area in the node and the edge between nodes. A reformulation of the algorithm is needed.

Furthermore, we also want to control the uniform distribution we gain from the Markov Chain, that's the reason why we use the Markov Chain method on the problem. Metropolis-Hastings Algorithm is needed to modify the transition matrix to in the process.